\documentclass[prd,
 reprint,
 amsmath,amssymb,
 aps,nofootinbib
]{revtex4-2}

\usepackage{bm}% bold math

\PassOptionsToPackage{linktocpage}{hyperref}
\usepackage[hyperindex,breaklinks,hidelinks]{hyperref}
\usepackage{enumitem}
\usepackage{slashed}
\usepackage[dvipsnames]{xcolor}

\usepackage{subcaption}

\usepackage{microtype}

\usepackage{tikz}
\usepackage[compat=1.1.0]{tikz-feynman}
\tikzset{every picture/.style={line width=0.75pt}} %set default line width to 0.75pt        
\usetikzlibrary{shapes.geometric}
\usetikzlibrary{decorations.markings}
\usepackage{orcidlink}
\usepackage{comment}
\usepackage{array}
\usepackage{mathtools}

\usepackage{etoolbox}

\begin{document} 

\title{Removing the Cosmological Bound on the Axion Scale \\ via Confinement During Inflation}

\author{Gia Dvali} 
\email{georgi.dvali@lmu.de}

\author{Sophia Fitz\,\orcidlink{0009-0001-6023-5640}}
\email{sophia.fitz@campus.lmu.de}

\author{Lucy Komisel\,\orcidlink{0009-0008-1578-588X}} 
\email{lucy.komisel@mpp.mpg.de}

\affiliation{Arnold Sommerfeld Center, Ludwig-Maximilians-Universit\"at, Theresienstra{\ss}e 37, 80333 M\"unchen, Germany}
\affiliation{Max-Planck-Institut f\"ur Physik, Boltzmannstr. 8, 85748 Garching, Germany}

\date{\today}

\begin{abstract} 

We implement the scenario of early relaxation of the axion via a high scale confinement within $SU(5)$ grand unified theory and study an epoch of strong QCD in inflationary cosmology. 
We consider scenarios in which, during inflation, the $SU(5)$ is either entirely or partially in the confining phase.
This generates an early potential for the axion and dilutes its energy density removing any cosmological upper bound on the decay constant.  
We show that a phase of strong QCD can be realized by at least two mechanisms:  
{\it 1)} a direct coupling between the inflaton and the gauge fields and/or 
{\it 2)} by restoration of the $SU(5)$ symmetry during the inflationary epoch.
In the latter case, strong coupling is already achieved via the RG running of the $SU(5)$ gauge coupling. 
We show that the mechanism works for all known realizations of the invisible axion idea: 
Peccei-Quinn (PQ) type formulations in which the anomalous global symmetry is realized via additional scalars (DFSZ) or heavy fermions (KSVZ) as well as the two-form gauge axion formulation based entirely on the QCD gauge redundancy without any anomalous global symmetry.
Even if the expectation value of the PQ scalar vanishes during inflation, the axion is a well defined degree of freedom represented by the phase of the fermion 't Hooft determinant. 
For the DFSZ case, this phase is composed out of a condensate of the ordinary quarks, amounting to an early universe version of the $\eta'$-meson.  
In all considered scenarios, the present day axion can be a viable dark matter candidate for an arbitrarily large value of the decay constant.      

\end{abstract}

\maketitle

\section{Introduction} 

This paper is about the axion's early relaxation mechanism via a period of early confinement, originally proposed in \cite{Dvali:1995ce}. 
This mechanism liberates the axion scale from any strict cosmological upper bound.
Our present goal is to provide an explicit implementation of this scenario within inflationary cosmology in a grand unified theory as well as to discuss some profound accompanying phenomena. 
Such is the breaking (both spontaneous as well as explicit) of the anomalous chiral symmetry by the condensate of the 't Hooft determinant, formed in the early epoch of high scale confinement. 
We start with a brief outline of some essentials of the axion physics.  

The QCD axion \cite{Weinberg:1977ma, Wilczek:1977pj} is a degree of freedom that makes the topological $\theta$-term, 
\begin{align}
    \theta \, {\rm tr} \, G_{\mu\nu}\tilde{G}^{\mu\nu} \,,
\end{align}  
unphysical. 
Here, $G_{\mu\nu}$ is the standard QCD field strength matrix and $\tilde{G}^{\mu\nu} \equiv \epsilon^{\mu\nu\alpha\beta}G_{\alpha\beta}$ is its Hodge dual.
In the absence of the axion, the physically measurable parameter, $\bar{\theta} \equiv  \theta + \mathrm{arg}(\det M_{Q}) $, is the combination of the ``bare" $\theta$ 
and the phase of the determinant of the quark mass matrix $M_Q$.
This parameter is measurable via the electric dipole moment of the neutron (nEDM) \cite{Baluni:1978rf,Crewther:1979pi}, resulting in the experimental upper bound $\bar{\theta}  \lesssim 10^{-9}$ \cite{Baker:2006ts, Pendlebury:2015lrz, Graner:2016ses}.
This is the source of the so-called strong-$CP$ puzzle.
The axion (if it has exact quality) forces the vacuum value $\bar{\theta} = 0$, thereby solving the puzzle.   
 
In the original Peccei-Quinn (PQ) formulation \cite{Peccei:1977hh, Peccei:1977ur}, the axion is introduced as a pseudo-Goldstone boson, $a(x)$, of a spontaneously broken anomalous global symmetry $U(1)_{PQ}$. 
That is, the axion describes phase fluctuations of the order parameter ${\mathcal O}(x)$ that breaks the anomalous global $U(1)_{PQ}$ symmetry spontaneously. 
Schematically, we can write,  
\begin{align} \label{Oa}
    {\mathcal O}(x) = |{\mathcal O}(x)| {\rm e}^{i\frac{a}{f_a}} \,.
\end{align}  
The operator ${\mathcal O}(x)$ can be an elementary field, a composite of several fields, or a combination of the two. 
Contributions to this order parameter can come from several sources. 
The axion is then defined as the proper superposition of the phases and its decay constant $f_a$ is determined accordingly.
Below the scale $f_a$, the $U(1)_{PQ}$ symmetry is realized non-linearly as a global shift symmetry, 
\begin{align} \label{aShift}
    a \rightarrow a + {\rm const}. 
\end{align}
The axion has exact quality, as long as the only source of explicit breaking of this symmetry is the QCD anomaly.  
This is not guaranteed in the PQ formulation, since the theory can be continuously deformed by operators that break the PQ symmetry explicitly. 

Alternatively, in a so-called gauge formulation \cite{Dvali:2005an}, the axion is introduced as an anti-symmetric $2$-form, $B_{\mu\nu}$, which transforms under the QCD gauge redundancy, without any reference to a global symmetry.  
Due to this, the gauge axion formulation has exact quality. 
In the low energy effective theory (EFT), the gauge formulation is dual (equivalent) to the PQ formulation, with exact quality. 
This is due to the QCD gauge symmetry of the gauge axion. 

However, despite the duality of the low-energy EFTs, the two formulations differ fundamentally in their UV theory above the scale $f_a$. 
Unlike the PQ axion, the gauge axion cannot be UV-completed into a complex scalar, and requires a UV-completion directly into a fundamental theory above the scale $f_a$ \cite{Dvali:2022fdv, Dvali:2023llt, Dvali:2024dlb}. 
Thus, in the gauge axion formulation, $f_a$ is the cut-off of the EFT.  
This can be viewed as the trade-off for the exact quality. 

Now, in axion physics there exist two important regime changes. 
The first one takes place when the axion becomes a valid degree of freedom. 
In the pseudo-Goldstone formulation, the axion emerges as a valid degree of freedom below the scale $f_a$, where the  chiral $U(1)_{PQ}$ invariance is broken spontaneously.     
For example, in standard  realizations of the PQ theory \cite{Peccei:1977hh,Peccei:1977ur}, in the current  cosmological epoch, the role of  ${\mathcal O}(x)$ is played by a complex scalar $\Phi = |\Phi(x)| {\rm e}^{\frac{a}{f_a}}$. 
The axion then represents its (pseudo)Goldstone phase, with the scale $f_a = \langle |\Phi | \rangle$ set by the vacuum expectation value (VEV) of the modulus \cite{Weinberg:1977ma, Wilczek:1977pj}.
However, we must stress that there is an admixture to the axion field from the phase of the fermion condensate.  
In the current universe, this admixture is negligible, since contributions to the spontaneous breaking of $U(1)_{PQ}$ from the quark condensate are very small compared to $f_a$.  
However, as we shall see later, this balance could have been very different in the early universe \cite{Dvali:2025xur}. 
  
In contrast, in the gauge axion case \cite{Dvali:2005an}, within the EFT, the axion is a valid degree of freedom at all times. 
Therefore, the scale $f_a$ must be above the relevant energy scales of all cosmological epochs describable within the EFT.     

The second important regime change takes place when the axion mass is generated.   
In the PQ formulation, this effect takes place at the scale where the shift symmetry \eqref{aShift} is explicitly broken by instanton effects.
That is, the axion potential acquires a tilt, giving the axion an effective mass $m_a$.   
Due to the equivalence of the two formulations in the low energy EFT, the same language can be used for the gauge axion case. 
  
In the context of high temperature QFT, these regime changes can be sequenced by the temperature of their corresponding phase transitions.  
In the commonly assumed case, the phase transition that breaks the PQ symmetry takes place at temperatures around the axion scale, $T \sim f_a$.  
Notice, however, that the temperature of the PQ phase transition is highly parameter dependent. 
In particular, as discussed in \cite{Dvali:1996zr}, for some values of the coupling constants, the PQ symmetry can remain broken at arbitrarily high temperatures due to the effect of high temperature symmetry non-restoration \cite{Weinberg:1974hy, Mohapatra:1979qt, Mohapatra:1979vr, Dvali:1995cc, Dvali:1995cj}.  

On the other hand, within the Standard Model (SM), the axion mass is generated below the QCD temperature $T_{\rm QCD} \sim 100\ \rm MeV.$ 
For higher temperatures, the instanton effects vanish, and the axion mass diminishes. 
The temperature $T_{\rm QCD}$ also marks the QCD phase transition, below which the theory enters the confining regime and the chiral quark condensate forms.  
    
Embedding the above thermal dynamics within standard hot big bang cosmology results in a mechanism of axion production via vacuum oscillations. 
After the axion mass is generated and overcomes the Hubble friction, the axion relaxes and starts coherent oscillations about the minimum.   
The axion energy density redshifts as matter and is highly sensitive to its initial amplitude.  
However, in this picture, prior to its mass generation, the axion potential has no preferred point. 
Correspondingly, within the minimalistic hot big bang cosmology, the amplitude cannot be predicted. 
One can therefore rely only on probabilistic and naturalness considerations.     
        
The commonly accepted cosmological bound \cite{Preskill:1982cy, Dine:1982ah, Abbott:1982af}, 
\begin{align}  \label{AxionBound}
     f_a \lesssim 10^{12}\  {\rm GeV}  ,
\end{align} 
was derived under the assumption that the initial amplitude of coherent oscillations at the QCD phase transition is maximal, $a_{in}/f_a \sim 1$, or equivalently, the misalignment angle is $\Delta \theta_{in} \sim 1$.   
However, as pointed out in \cite{Dvali:1995ce}, this assumption is too strong as it assumed that the axion had no prior knowledge of the minimum of its potential before the (latest) QCD phase transition.    
As was shown there, this assumption is invalid in a large class of cosmological scenarios. 
In particular, this is the case for inflationary cosmology. 
    
The point of \cite{Dvali:1995ce} is that during inflation the strength of the QCD gauge coupling
as well as the axion decay constant should have been significantly different from its value in the present epoch, due to very large displacements of the fields from their current values. 
Such displacements are an intrinsic part of the inflationary cosmology. After all, the very concept of inflation is based on the premise  that fields change in time. 

Due to this, the QCD coupling, which is field dependent, could have become strong, resulting in an early epoch during which the mass of the axion field exceeds the value of the Hubble parameter. 
During this period, the axion undergoes a series of damped coherent oscillations, quickly relaxing the misalignment angle $\theta_{\rm eff}$.
A short epoch of relaxation suffices to lift the upper bound on today's value of the axion scale $f_a$. 
Some further studies and implementations of the early relaxation scenario \cite{Dvali:1995ce} can be found in the following (incomplete) set of references \cite{Choi:1996fs, Banks:1996ea, Takahashi:2015waa, Tokareva:2017nng, Co:2018mho, Co:2018phi, Berger:2019yxb, Buen-Abad:2019uoc, Matsui:2020wfx, DvaliGomez, Reig:2021ipa, Koutsangelas:2022lte, Ettengruber:2023tac, Kitano:2023mra}.

In the present article we are going to highlight a number of important factors characteristic to the early relaxation mechanism that were not entirely appreciated previously. 
     
In particular, during the cosmological evolution not only the mass and the decay constant, but the very nature of the axion degree of freedom is generically time-dependent.
That is, the superposition and the types of fields that cancel the vacuum $\bar{\theta}$-angle change throughout the Universes early history. 
For example, a vanishing of the PQ field VEV $\langle \Phi \rangle $ in a strong QCD epoch by no means implies the restoration of the PQ symmetry. 
Instead the symmetry is broken by the condensate of the corresponding fermionic 't Hooft determinant, with the axion represented by its phase.

Notice that this behavior takes place in a very wide range of the parameter space and, therefore, must be taken into account whenever one talks about the axion cosmology.   
     
In the present article, we demonstrate the above features by studying the cosmology of the axion in a simple inflationary scenario with $SU(5)$ grand unification.
We show that a phase of strong QCD in the early epoch can result from various factors.
First, the QCD gauge coupling can have a direct field dependence due to its couplings with grand unified and inflaton fields.  
Secondly, a high scale QCD confinement in the early epoch can be a result of a non-thermal restoration of the grand unified $SU(5)$-symmetry.  

As said, we show  that the constituency of the axion can significantly vary over time.
In particular, in an early epoch the main contribution may come from the phase of the condensate of 't Hooft's quark determinant. 
In case of the grand unified symmetry, the condensate forms due to $SU(5)$ instantons. 
In all scenarios, the early oscillations of the axion can reduce the axion energy density to an acceptable value for arbitrarily high $f_a$, essentially abolishing the upper bound \eqref{AxionBound} on the axion scale. 
    
\section{The Setup} 
  
The gauge sector of the theory is described by the $SU(5)$ vector bosons in the adjoint representation. 
The corresponding field strength matrix shall be denoted by $G_{\mu\nu}$.
   
The key participants of the scalar sector of the theory are the following. 
First, there is the PQ field $\Phi$ and the inflaton field $S$. 
Both are assumed to be singlets under the $SU(5)$-symmetry.
The inflaton $S$ is taken to be a $CP$-even real scalar. This is required in order for the inflaton  not to generate an additional $CP$-violating contribution to $\bar{\theta}$, apart from standard renormalization running of $\mathrm{arg}(\det M_{Q})$.

In addition, we have the Higgs sector of the theory that includes the $24$-dimensional adjoint field $\Sigma_i^j$ and at least one field $H_i$ transforming in the $5$-dimensional fundamental representation. 
Here and below, $i,j=1,2,...,5$ denote the $SU(5)$ indices.     
The fermion sector of the theory shall be specified according to a concrete realization of the PQ mechanism.  

Now, the master field is the inflaton field $S$, as it determines the phase portrait of the theory. 
Through its couplings, the value of $S$ controls the expectation values of $\Phi, \Sigma, H$ as well as the value of the gauge coupling. 
  
The gauge coupling $\alpha_{\rm eff}$ effectively becomes a function of $S$. 
This phenomenon can be described via an effective term in the Lagrangian, 
\begin{align} \label{AlphaS}
    \propto\alpha_{\rm eff}^{-1} (S/M) {\rm tr} G_{\mu\nu} G^{\mu\nu} \,,
\end{align}  
where $M$ is a scale  that depends on the origin of the above effective operator. This can come 
from the following sources:

1) Higher dimensional operators emerging at the cutoff scale $M_{UV}$; 
   
2) Perturbative renormalization effects due to the dependence of the Higgs VEVs on the inflaton field. In this case, the scale $M$ is set by the masses of the heavy fields.
This source is calculable within the EFT.  
  
In addition, the scale of spontaneous breaking of the PQ symmetry is also controlled by the inflaton field $S$.  
Notice that the breaking of $U(1)_{PQ}$ has two contributions:  
  
1) The VEV of the PQ field $\Phi$, which depends on the inflaton field $S$;
  
2) The condensate of chiral fermions transforming under $U(1)_{PQ}$. 
The universal order parameter is the VEV of the 't Hooft determinant. 
For massless fermions the scale of the condensate is given by the QCD scale, $\Lambda_C$, which is also a function of $S$, as discussed above. 

Due to this, not only the axion mass, but also its composition is $S$-dependent.
For any given value of $S$, it corresponds to a superposition of the (properly normalized) phases of $\Phi$ and of the 't Hooft determinant. 
The specific form of this composition is model-dependent and will be discussed in more detail later (see, also the Appendix B). 

Let us now discuss how the effective mass of the axion, $m_a(S)$, can become a function of the inflaton field $S$. 
Let us denote the canonically normalized contemporary axion field by $a$ and its decay constant by $f_a$. 
    
Then, the equation of motion for $a$ in the inflationary universe is given by 
\begin{align} \label{axionEq}
    \ddot{a} + 3 {\mathcal H} \, \dot{a} \,  + \, \Lambda_C^4 
    \frac{\partial}{\partial a} {\mathcal V}\left ( \frac{a}{f_a} -
   \bar{\theta} \right ) = 0 \,,  
\end{align}
where $\mathcal H$ is the Hubble parameter and ${\mathcal V}$ is a periodic function of its argument, which represents the effective vacuum angle $\theta_{\rm eff}(a)  \equiv \frac{a}{f_a} -\bar{\theta}$. 
The minimum of this function is strictly at the zero value of the argument, i.e.,
\begin{align} \label{mintheta}
    \theta_{\rm eff} \, \equiv \, \frac{a}{f_a} -
   \bar{\theta} \, = \, 0\,.
\end{align}    
This is an exact statement and is independent of the identity of the axion field \cite{Vafa:1984xg}. In fact, it has been shown that not only the global minimum, but any extremum of the function  ${\mathcal V}$ is $CP$-conserving \cite{Dvali:2005an}.

The precise form of the function ${\mathcal V}(\theta_{\rm eff})$ is not known.     
In the dilute instanton gas approximation, it has the well known form     
${\mathcal V} = - \cos \left ( \frac{a}{f_a} - \bar{\theta} \right )$, which corresponds to the axion potential     
\begin{align}
    V(a) =  \Lambda^4_C {\mathcal V}(a)  = 
    - \Lambda_C^4 \, \cos \left ( \frac{a}{f_a} - \bar{\theta} \right)\,.
\end{align}   
Of course, an overall constant must be added to the scalar potential in order to fix the vacuum energy (close) to $0$.
However, the precise form of ${\mathcal V}(\theta_{\rm eff})$ is unimportant for knowing that the exact minimum is always at $\theta_{\rm eff} = 0$.      
The axion potential is generated by instanton effects and is the only term in the effective Lagrangian that breaks the anomalous shift symmetry,
\begin{align} \label{shiftA}
   \frac{a}{f_a} \rightarrow \frac{a}{f_a} + {\rm const.} 
\end{align}  
As mentioned previously, all parameters entering eqn. \eqref{axionEq} (${\mathcal H}\,, \Lambda_C\,, f_a$) are functions of the inflaton field $S$.  
The composition of the axion field is also time dependent via $S$. 
However, at any time, the Lagrangian maintains the shift symmetry \eqref{shiftA} modulo the contribution from the anomaly. 
Due to this, w.l.o.g., we can absorb $\bar{\theta}$ by shifting the axion field, so that the minimum is at $a=0$. 
The axion mass is then given by the curvature of the potential around the minimum, 
\begin{align} 
    m_a^2 \, \equiv\Lambda_C^4 \left. \frac{\partial^2 {\mathcal V}}{\partial a^2}
    \right|_{a=0}\simeq \frac{\Lambda_C^4}{f_a^2}\,.
\end{align} 
Then, up to nonlinearities, eqn. \eqref{axionEq} can be expanded in $a$ as 
\begin{align} \label{axionEq2}
    \ddot{a} + 3 {\mathcal H} \, \dot{a} \,  + \, m_a^2 a\, + \, .. = 0 \,.  
\end{align}
Non-linear terms become important only for a large amplitude $a \sim f_a$. However, even then they do not change the picture qualitatively.  
    
Now, the key point of the early relaxation scenario is that during a certain period in the inflationary epoch, marked by some initial $S_{i}$ and final $S_{f}$ values of the inflaton field, the axion mass $m_a$ exceeds the value of the inflationary Hubble parameter.   
Then, for $S  \in  [S_{i} ,  S_{f}]$, the axion undergoes damped oscillations, relaxing the vacuum angle $\theta_{\rm eff}$ to zero. 
In this way, the misalignment angle is drastically reduced already during inflation.   
The axion amplitude decreases as, 
\begin{align} 
    \label{deltaA}
    a_f = a_i {\rm e}^{-\frac{3}{2} N_e} \,, 
\end{align} 
where $N_e \equiv  {\mathcal H}t$ is the number of inflationary e-foldings during this period. 
Correspondingly, the vacuum angle gets reduced as $\theta_{f} = \theta_{i} {\rm e}^{-\frac{3}{2} N_e}$.  

It is reasonable to expect that after inflation and reheating, the universe passes through an epoch in which the temperature $T$ exceeds the contemporary value of $\Lambda_C$. 
Then, the axion potential flattens and remains so until $\Lambda_C > T$ is satisfied again. 
Of course, a temporarily flattening and subsequent tilting of the axion potential could have happened more than once. 
The latest iteration took place after the ``ordinary" (i.e., the last)  QCD phase transition. 
   
No matter how many times the flattening of the axion potential takes place,
as long as the axion remains a valid degree of freedom,
the memory about the misalignment angle is maintained. 
In the PQ formulation this is guaranteed as long as the global PQ symmetry remains spontaneously broken, whereas in the gauge axion case this condition is satisfied automatically. 

In such cases, after the regular QCD phase transition, the axion starts its oscillations with a reduced amplitude,
\begin{align} \label{fsmall} 
    a_0 \lesssim  f_a  {\rm e}^{-\frac{3}{2} N_e} \,.
\end{align}     
This correspondingly reduces the present day axion density by a factor $\sim {\rm e}^{- 3 N_e}$, as compared to the case with a maximal misalignment angle. 
Even for a moderate number of e-foldings with a strong QCD phase, say $N_e \gtrsim 10$, this completely lifts any cosmological constraint  on the axion decay constant.       
    
\section{The Inflationary Shift Mechanism} 
    
Let us now discuss the prototype potential that leads to the shift of the VEVs during inflation.  
The key ingredient is the interaction between the axion and other fields, which we shall group together with the mass terms as
\begin{align}
\begin{split}
    V_S =   
   &(\kappa_\Sigma S^2- M_{\Sigma}^2 ) \mathrm{tr}\Sigma^2  
   + (\kappa_\Phi S^2 - M_{\Phi}^2) \Phi^\dagger \Phi \\
   +&(\kappa_H S^2 - M_{H}^2) H^\dagger H   \,,  
\end{split}
\end{align}
where the parameters $\kappa_\Sigma > 0$, $\kappa_\Phi > 0$, $\kappa_H > 0$ are real and positive definite.  
The above interactions make the mass terms of the fields dependent on $S$:
\begin{align} \label{MassesSD}
\begin{split}
M_{\Sigma}^2(S) &\equiv \kappa_\Sigma S^2- M_{\Sigma}^2 \,,\\
M_{H}^2(S) & \equiv  \kappa_H S^2 - M_{H}^2\,, \\
M_{\Phi}^2(S) & \equiv \kappa_\Phi S^2 - M_{\Phi}^2 \,.
\end{split}
\end{align}
  
The rest of the potential consists of two parts. 
The first one, which we denote by $V_{\Sigma, H, \Phi}$, contains all possible $SU(5)\times U(1)_{PQ}$-invariant interaction terms 
of $\Sigma, H, \Phi$.  
E.g., considering only renormalizable interactions gives  
\begin{align} \label{scalarpot}
\begin{split}
    V_{\Sigma, H, \Phi} &=  \lambda_\Sigma \mathrm{tr}\Sigma^4 + \lambda_\Sigma' (\mathrm{tr}\Sigma^2)^2 + \mu_{\Sigma} \mathrm{tr} \Sigma^3  \\
    &+\lambda_H (H^\dagger H)^2 + \lambda_\Phi (\Phi^\dagger \Phi)^2 \\
    &-\alpha_\Sigma (\mathrm{tr}\Sigma^2) \Phi^\dagger \Phi+ \alpha_H H^\dagger H \Phi^\dagger \Phi \\
    &+ \lambda_H' (\mathrm{tr} \Sigma^2) H^\dagger H +\beta_H H^\dagger \Sigma^2 H + \mu_{H}  H^\dagger \Sigma H \\
    &+ V_0 ,
\end{split}
\end{align}
where the first two lines contain the self-interaction terms, the third and fourth lines contain all possible cross-couplings and the last line is an overall constant that tunes the energy of the global minimum to the present value of the vacuum energy. 

The second part of the potential is the self-interaction of the inflaton field $V_S$.  
This potential must satisfy the slow roll conditions in a certain interval of $S$.
Other than this, the explicit form is unimportant for our considerations. 
W.l.o.g., we assume that in the global minimum (the vacuum today), the expectation value of the inflaton $\langle S\rangle$ is zero.  
 
For values of the inflaton field $S \gg \frac{M_{\Sigma}}{\sqrt{\kappa_{\Sigma}}}, \frac{M_{H}}{\sqrt{\kappa_{H}}},  \frac{M_{\Phi}}{\sqrt{\kappa_{\Phi}}} $  
the effective mass terms of all three fields are positive, and their expectation values vanish.  
The inflationary potential can be reduced to, 
\begin{align} \label{Vinf}
   V = V_0 + V(S).
\end{align}   
We assume that $V(S) \ll V_0$, so that the inflationary universe is dominated by the (almost) constant vacuum energy $V_0$ as for example in the scenarios of  \cite{Linde:1993cn, Copeland:1994vg, Dvali:1994ms, Binetruy:1996xj}.  
  
In the minimalistic case we can take simply $V(S) =0$, which will make the tree-level inflaton potential exactly flat.  
However, as shown in \cite{Dvali:1994ms, Binetruy:1996xj}, corrections are already generated at one-loop via the Coleman-Weinberg effective potential. 
This correction cannot be tuned away, even in supersymmetric scenarios, and they have the asymptotic form (as $S$ is much larger than the ``bare" masses in (\ref{MassesSD})) \cite{Dvali:1994ms, Binetruy:1996xj} 
\begin{align}  \label{Fterm} 
   V(S) \propto V_0 \ln(S)\,. 
\end{align}      
However, the above is just one elegant example and our scenario is in principle open to any suitable form of the inflaton potential
\footnote{We notice that, if non-renormalizable couplings of the type \eqref{AlphaS} are present, they can generate additional contributions to the inflaton potential.  
These contributions do not necessarily ruin inflation and can in fact serve as its primary source, as explicitly shown in \cite{Dimopoulos:1997fv}.}.  

The important thing for us is that during a certain stage of inflation the expectation values of the PQ field and possibly the GUT Higgs fields vanish. 
   
If the GUT fields vanish, the GUT symmetry is fully restored in this epoch.  
In contrast, the story with the PQ symmetry is more subtle. 
Even though the PQ field vanishes, the PQ symmetry is broken by the fermion condensate. 
This will be analysed in detail later.  
   
\section{Strong QCD in the Inflationary Epoch} 
   
The next ingredient of our scenario is the behaviour of the QCD gauge coupling $\alpha_{\rm QCD}$ during the inflationary epoch.  
This coupling is affected, the least, by the following two sources. 
First we notice that, what a low energy observer measures in the current universe as the QCD coupling, is the inverse coefficient of the gluon field strength.
This coefficient is field dependent. 
This dependence comes from two sources which we shall consider separately. 

\subsection{Cutoff Suppressed Operators}

The first source are high-dimensional operators, suppressed by the theory's cutoff, $M_{UV}$. 
Since gluons are part of the $SU(5)$ gauge sector, we simply use the $SU(5)$ gauge field strength, 
\begin{align} \label{GaugeKin}
   {\rm tr} \left({\mathcal Q}(S,\Sigma, H) \, G_{\mu\nu}G^{\mu\nu} \right),
\end{align}
where  ${\mathcal Q}(S,\Sigma, H)$ is a generic gauge kinetic function of the scalar fields 
that transforms as the reducible $24 \times 24$ tensor product under $SU(5)$.  
The above expression can contain an infinite series of invariants.  
Some examples of dim-5 and dim-6 level operators are: 
\begin{align}
\begin{split}
    &S  {\rm tr} (G_{\mu\nu}G^{\mu\nu})  +  {\rm tr} ( \Sigma \, G_{\mu\nu}G^{\mu\nu})   \\      
    &S^2   {\rm tr} (G_{\mu\nu}G^{\mu\nu})  +  {\rm tr} (\Sigma^2 G_{\mu\nu}G^{\mu\nu})  + 
    {\rm tr} (H^{\dagger}H G_{\mu\nu}G^{\mu\nu})\, + \,...
\end{split}
\end{align}
Of course, these are suppressed by the proper powers of the cut-off scale $M_{UV}$.   
What we measure as the gauge coupling today is the net effect  of all the above operators in which the scalar fields are replaced by their VEVs. 
However, as we showed above, these VEVs  during inflation were very different from their current values. 
Correspondingly, in that epoch the effective gauge couplings were dramatically different from what we measure today. 
        
In the absence of a microscopic theory, we must treat the above gauge kinetic function as generic. 
Correspondingly, for a very large portion of the theory space, the QCD coupling becomes strong during some period of inflation.  

Regarding the cutoff $M_{UV}$, it is important to understand the following two things. 
First, the cutoff $M_{UV}$ is always below the Planck mass, $M_P$. 
In particular, according to non-perturbative gravitational arguments \cite{Dvali:2007hz, Dvali:2007wp, Dvali:2008fd, Dvali:2008ec}, the cutoff is necessarily bounded from above by the species scale  $M_{UV} \leqslant M_P/\sqrt{N}$, where $N$ is the number of particle species. 
In all models of interest, this number is above $100$. 
Correspondingly, due to this reasoning alone, the cutoff $M_{UV}$ is not far from the GUT scale.   

Secondly, the time-variations of the inflaton field are not restricted by the cutoff scale $M_{UV}$.  Instead, the validity of the inflationary scenario is strictly constrained by the upper bound on the Hubble scale  ${\mathcal H} \lesssim M_{UV}$.  
In the following, it will be assumed that this constraint is fully satisfied.     

\subsection{Inflaton-Dependent Renormalization Effects}
  
The second source of strong QCD in the inflationary epoch exists regardless of the high dimensional operators generated by fundamental physics above $M_{UV}$. 
This source is the restoration of the $SU(5)$ symmetry. 
As already discussed, the restoration of the $SU(5)$ symmetry is possible since, for large values of the inflaton $S$, the couplings \eqref{MassesSD} generate large positive mass terms for the GUT Higgs fields. 
Correspondingly, the scalar VEVs vanish, $\langle H\rangle = \langle \Sigma \rangle = 0$, such that all SM fermions are massless.
Under such conditions, the running gauge coupling of the restored $SU(5)$ can become strong at the scale $\Lambda_C \sim 10^7$ GeV, depending on the specific realization of the axion. 
A more detailed discussion of this regime for different axion models is given in  Appendix \ref{AppA}.

This exceeds the  QCD scale of the $SU(3)\times SU(2) \times U(1)$ theory obtained in RG running under identical boundary conditions at the GUT scale. 
Correspondingly,  with the $SU(5)$-symmetry restored during inflation, QCD can become strong, as a part of $SU(5)$ within a purely renormalizable framework. 

Notice that the change of the effective gauge coupling by the alteration of the RG running can also be understood in terms of an effective coupling between the inflaton and the gauge fields obtained by integrating out the $SU(5)$ fields with inflaton-dependent masses.

In what follows, we shall assume that the gauge coupling becomes strong due to one (or both) of the above two effects, and that the corresponding QCD scale $\Lambda_C$ exceeds the contemporary Hubble parameter. 
      
Notice that inflationary scenarios with Hubble parameter ${\mathcal H} \sim 10^5$GeV are perfectly viable, since the reheating temperature $T_R \sim \sqrt{{\mathcal H} M_P}$ is sufficiently high both for consistent baryogenesis as well as for the generation of density perturbations via the mechanism of modulated reheating \cite{Dvali:2003em, Dvali:2003ar}.       

\subsection{Axion During Inflation}
      
We shall now discuss the fate of the axion during inflation.   
We first outline generic features and will later enter into a model-dependent analysis. 
        
It is very important to understand that a vanishing VEV of the PQ field $\Phi$ does not imply the restoration of the PQ symmetry.   
Instead, since QCD is in the strong phase, the PQ symmetry is spontaneously broken by the chiral fermion condensate, as at least one quark must transform under the PQ symmetry non-trivially.  
Namely, the relevant order parameter is the 't Hooft determinant of those fermions that transform  under $U(1)_{PQ}$.           
The precise structure of this determinant is model dependent and will be deconstructed later for concrete scenarios. Here we only outline the generic features. 

We schematically  denote the relevant part of the 't Hooft determinant by  $det{\psi}$. 
Of course, this structure is gauge-invariant and the number of fermions entering in it is given by the number of the fermionic zero modes existing in the instanton background. 
The latter  depends on the fermion representation content.

Notice that the non-zero VEV of the 't Hooft determinant is a universal feature of theories with chiral fermions, which directly follows from the topological susceptibility of the vacuum for the corresponding gauge theory and its removal by the chiral symmetry of fermions. 
This necessitates the existence of a phase degree of freedom, coming from the fermion condensate. 
For a detailed discussion the reader is referred to \cite{Dvali:2017mpy}.

The non-zero VEV of $det{\psi}$, breaks the $U(1)_{PQ}$ spontaneously. 
The corresponding phase degree of freedom, which we denote by $\eta_{F}$  plays the role of the axion.  
That is, in the regime in which $\langle\Phi\rangle$ vanishes, the axion is entirely given by $\eta_{F}$. 
If $\Phi$ maintains a VEV, the axion is a combination of the two phases, weighted by their respective VEVs. 
     
Notice that the existence of the condensate does not require the fermions to be massless, but only that their masses and Yukawa couplings respect the anomalous $U(1)_{PQ}$-symmetry. 
This ensures the fermionic zero mode structure of the instantons, which triggers the condensate \cite{Dvali:2024zpc, Dvali:2025pcx}.

\section{A KSVZ Model}    
  
In the case of a Kim-Shifman-Vainshtein-Zakharov type (KSVZ) scenario \cite{Kim:1979if, Shifman:1979if}, the Yukawa couplings of the ordinary SM fermions do not exhibit any global symmetry that can serve as $U(1)_{PQ}$-symmetry. 
Therefore, the fermions transforming under $U(1)_{PQ}$ must be introduced as external heavy fermions. 
We shall now embed this model into our $SU(5)$-framework. 
 
As usual, each generation of quarks and leptons is introduced as left-handed Weyl fermions transforming under the $SU(5)$-symmetry in the anti-symmetric $\mathbf{10}$ and anti-fundamental $\mathbf{\bar{5}}$ representations. 
For simplicity, we only consider a single generation and denote the fermionic fields as $\mathbf{10}_{jk}$ and  $\mathbf{\Bar{5}}^j$, respectively.
We introduce the additional heavy quarks as left-handed Weyl fermions in the fundamental and anti-fundamental representations, 
denoted by $\Bar{\Psi}^i$ and $\Psi_i$ respectively ($\bar \Psi^i$ does not refer to the Dirac-conjugate but to a different Weyl fermion in the anti-fundamental representation of $SU(5)$).  
The Yukawa couplings have the form  
\begin{align}
\begin{split}
    \mathcal{L}_{\rm Y} &= g^u H_i(\mathbf{10}_{jk}\mathbf{10}_{mn}\epsilon^{ijkmn}) + g^d H^{\dagger i}(\mathbf{10}_{ij}\mathbf{\Bar{5}}^j) \\ 
    &+ g^F \Phi \Bar{\Psi}^i\Psi_i + {\rm h.c.} \,,
\end{split}
\end{align}
where the Lorentz-spinor indices and charge conjugation matrices are not shown explicitly.  

The Yukawa couplings of the ordinary SM fermions ($\mathbf{10}_{jk}$ and  $\mathbf{\Bar{5}}^j$) exhibit no global chiral symmetry. 
Only the heavy fermions $\Bar{\Psi}^i\,, \Psi_i$, which directly couple to the PQ field $\Phi$, transform under the anomalous $\mathrm{U}(1)_\mathrm{PQ}$-symmetry as, 
\begin{align}
\begin{split}
    &\Bar{\Psi}\rightarrow e^{i\frac{\alpha}{2}} \Bar{\Psi}, \quad \Psi\rightarrow e^{i\frac{\alpha}{2}} \Psi\,, \\
   &\Phi\rightarrow e^{-i\alpha}\Phi\,.
\end{split}
\end{align}
In the standard treatment, the cosmology of this scenario works as follows.  
It is assumed that the axion potential is generated only when the temperature drops below the ordinary QCD scale $T \sim \Lambda_C \sim 100\ \rm MeV$. 
By then, the GUT symmetry is already Higgsed and the PQ-symmetry is spontaneously broken by the VEV of $\Phi$ at the scale $f_a \gg \Lambda_C$. 
The axion then is represented by the phase of $\Phi$ with a tiny admixture from the phase of the fermionic 't Hooft determinant.  
Since under a standard assumption the axion has no prior knowledge of its minimum, it is natural to assume that the misalignment angle $\theta_{\rm eff}$ is maximal. 
That is, the initial amplitude is $a_0 \sim f_a$. 
This results in the cosmological upper bound \eqref{AxionBound}.
   
However, as we have argued, with a high likelihood, the real history could have been very different. 
With the above fermionic content, let us consider an inflationary epoch during which the effective masses of $\Phi$ and the GUT fields are positive, and QCD dynamics is strongly coupled.  
 
In the absence of strongly coupled gauge dynamics, the PQ-symmetry would be restored.  
However, the QCD scale dominates over the Hubble scale.   
In this case, the effects of the space-time curvature and of the de Sitter temperature can be ignored and the instantons operate as they would in zero-temperature flat space.  
    
Correspondingly, the chiral quark condensate forms, spontaneously breaking the flavor symmetries.  
We are mainly interested in the breaking of the $U(1)_{PQ}$. 
The order parameter of this breaking is the 't Hooft determinant of the chiral fermions. 
Notice that although all the fermion flavors enter, the important part is the contribution to the condensate from $\Bar{\Psi}^i\Psi_i$, which  transforms non-trivially under the PQ symmetry. 
The relevant order parameter is  $\langle  \Bar{\Psi}^i\Psi_i \rangle $.  
This condensate is non-zero and is of order $\sim \Lambda_C^3$. 
     
\begin{figure}
    \centering
    \begin{tikzpicture}
    \def\r{1}
    \begin{feynman}
        \vertex[blob] (t) at (0,0){};
        \vertex (pq1) at (-0.75*\r,1.5*0.866*\r){$\Psi$};
        \vertex (pq2) at (-0.75*\r,-1.5*0.866*\r){$\bar \Psi$};
        \vertex (5) at (-1.5*\r,0){$\bar{\mathbf{5}}$};
        \vertex (101) at (0.75*\r,1.5*0.866*\r){$\mathbf{10}$};
        \vertex (102) at (0.75*\r,-1.5*0.866*\r){$\mathbf{10}$};
        \vertex (103) at (1.5*\r,0){$\mathbf{10}$};

        \diagram{
            (5) --[fermion] (t),
            (pq1) --[fermion] (t),
            (pq2) --[fermion] (t),
            (101) --[fermion] (t),
            (102) --[fermion] (t),
            (103) --[fermion] (t)};
    \end{feynman}
    \end{tikzpicture}
    \caption{A diagrammatic representation of the 't Hooft vertex in the KSVZ-like model. The 
    fundamental ($\mathbf 5$) and anti-fundamental $\bar{\mathbf 5}$ fermions  each contribute one zero mode in the instanton background, whereas the fermion in the antisymmetric $\mathbf 10$ contributes three.}
    \label{fig:KSVZdeterminant}
\end{figure}
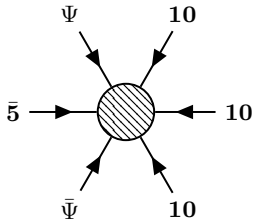

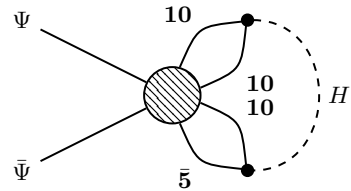
\begin{figure}
    \centering
    \begin{tikzpicture}
    \def\r{1}
    \begin{feynman}
        \vertex[blob] (t) at (2*\r,0){};
        \vertex (i1) at (0, \r){$\Psi$};
        \vertex (i2) at (0, -\r){$\bar \Psi$};
        \vertex[dot] (x1) at (3*\r,\r) {};
        \vertex[dot] (x2) at (3*\r,-\r){};

        \diagram{
            (i1) --[solid] (t),
            (i2) --[solid] (t),
            (x1) --[solid, half right, out = 30, in = 150, edge label = \(\mathbf{10}\) ] (t) --[solid, half right, in = 150, out = 30, edge label = \(\mathbf{10}\)] (x1),
            (x2) --[solid, half right, out = 30, in = 150, edge label = \(\bar{\mathbf{5}}\)] (t) --[solid, half right,  out = 30, in = 150, edge label = \(\mathbf{10}\)] (x2),
            (x1) --[scalar, half left, edge label = \(H\)] (x2)};
    \end{feynman}
    \end{tikzpicture}
    \caption{A diagrammatic representation of the dressing of the 't Hooft vertex by the exchange of a virtual Higgs.}
    \label{fig:dresseddeterminant}
\end{figure}

Notice that, even though the VEV of the SM Higgs $H$ vanishes and the $\mathbf{10}$ and $\mathbf{\Bar{5}}$ fermions are massless, there is no chiral symmetry. 
Schematically, the complete 't Hooft determinant 
consists of the gauge invariant terms of the following type (see also Fig. \ref{fig:KSVZdeterminant}),
\begin{align}
    \frac{1}{\Lambda^5} \, (\Bar{\Psi}^r\Psi_{r} )(\mathbf{10}_{jk}\mathbf{10}_{mn}\mathbf{10}_{il}\mathbf{\Bar{5}}^l \epsilon^{ijkmn}) \,+\,...\,.
\end{align}
We have singled out the structure in which PQ fermions form an autonomous $SU(5)$ invariant  $\Bar{\Psi}^r\Psi_{r}$.  
This is because the exchange of a virtual SM Higgs (see the diagram in Fig. \ref{fig:dresseddeterminant}) generates another effective operator, 
\begin{align}
    \Lambda_C (\Bar{\Psi}^r\Psi_{r} ),
\end{align}
which makes the contribution of ordinary quarks and leptons less relevant.

\begin{figure}
    \centering
    \begin{tikzpicture}
    \def\r{1}
    \begin{feynman}
        \vertex (t) at (0,0);
        \vertex[crossed dot] (i1) at (\r,0){};
        \vertex (l) at (1.6*\r,0)[xshift=12pt]{$\langle\bar\Psi\Psi\rangle$};
        \vertex (i2) at (-\r,0){$\Phi$};

        \diagram{
            (i2) --[scalar] (t),
            (t) --[fermion, half right, out=80, in=100] (i1) --[anti fermion, half right, out=80, in =100] (t),};
    \end{feynman}
    \end{tikzpicture}
    \caption{A Feynman diagram contributing to the tadpole for  $\Phi$.}
    \label{fig:tadpolecondensate}
\end{figure}
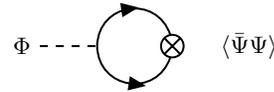

Notice that the fermion condensate also creates a tadpole for $\Phi$ (see also Fig. \ref{fig:tadpolecondensate}), 
\begin{align}
    g^F \Phi \langle \Bar{\Psi}^i\Psi_{i} \rangle,
\end{align}
which induces a VEV of $\Phi$ of the order,   
\begin{align} \label{PhiVEV}
|\langle \Phi \rangle| \sim  \frac{1}{M^2_\Phi(S)} g^F \langle \Bar{\Psi}^i\Psi_{i} \rangle  \sim \Lambda_C^3/M_{GUT}^2.
\end{align}

All parameters are evaluated for relevant values of the inflaton field, which typically is  $S \gg M_{GUT}$.  
Without a specific fine-tunning, the masses are also very large. In particular, it is expected that  $M_{\Phi} \gg \Lambda_C$.  
In such a case, the dominant contribution to PQ breaking is coming from the fermion condensate. 
Correspondingly, the axion is predominantly made up of the phase of the condensate $\langle \Bar{\Psi}^r\Psi_{r} \rangle $.  
This phase plays the role of the $\eta'$-meson for the inflationary epoch.
We can call it $\eta_{F}$. 
This phase will very quickly relax the effective vacuum angle $\theta_{\rm eff}$ to its minimum.
More details can be found in Appendix B. 

At the end of this section, we would like to comment that the generalization of the above analysis for the case when a period of strong QCD overlaps with broken GUT symmetry is straightforward. 
For example, if the $SU(3)\times SU(2)\times U(1)$-invariant VEV of the $\Sigma$-field is much greater than the strong coupling scale of QCD, $\Lambda_C$, the fermion condensate is generated predominantly by the $SU(3)$ and $SU(2)$ instantons. 
The contributions from the constrained instantons involving the superheavy GUT gauge fields are strongly  suppressed.  
Correspondingly, to identify the relevant order parameters, one can focus on the $SU(3)\times SU(2) \times U(1)$-invariant reduction of the 't Hooft determinant.   
This applies to all realizations of our scenario in the invisible axion cosmology.

\section{A DFSZ Model}
Let us now discuss a Dine-Fischler-Srednicki-Zhitnitsky (DFSZ) type invisible axion scenario \cite{Dine:1981rt, Zhitnitsky:1980tq}. 
In this version, no fermions beyond the SM ones are introduced. 
The $U(1)_{PQ}$-symmetry is realised by ordinary quarks and leptons. 
In order to allow for such a symmetry one needs to introduce a second Higgs doublet. 
In the present case, this doublet is a part of an additional Higgs $5$-plet.  
The two Higgs $5$-plets give masses to the separate (up and down) fermion flavors. 
We  therefore denote them by $H_u$ and $H_d$ respectively. 
The Yukawa couplings have the form, 
\begin{align}
    \mathcal{L}_{Y} = g^uH_{u,i}\mathbf{10}_{jk}\mathbf{10}_{mn}\epsilon^{ijkmn} + g^d \mathbf{\Bar{5}}^{j}\mathbf{10}_{ij}H_d^{\dagger i} + \mathrm{h.c.}\,.
\end{align}
These Yukawa couplings are invariant under the following chiral $\mathrm{U}(1)_{PQ}$-symmetry,
\begin{align} \label{DFSZsymtrafo}
\begin{split}
     &H_u \rightarrow e^{-i\alpha}H_u, \qquad H_d\rightarrow e^{i\alpha}H_d , \\
     &\mathbf{10}\rightarrow e^{i\frac{\alpha}{2}}\mathbf{10}, \qquad \mathbf{\Bar{5}}\rightarrow e^{i\frac{\alpha}{2}}\mathbf{\Bar{5}} \,.
\end{split}
\end{align}
The relevant new term in the scalar potential is the coupling between the  Higgs $5$-plets and the 
PQ field, which (at the renormalizable level) can be introduced in one of the following two forms, 
\begin{align} \label{HuHdPP}
    \mu_{\Phi} H_u^{\dagger} H_d \Phi^\dagger\, ~ {\rm or} ~
    \,  \xi_{\Phi}  H_u^{\dagger}  H_d \Phi^\dagger \Phi^\dagger + \mathrm{h.c.} \,,
\end{align}
where $\mu_{\Phi}, \xi_{\Phi}$ are parameters. 
These  terms fix the PQ transformation properties of the  $\Phi$-field as, 
\begin{align}\label{DFSZsymtrafo2}
    \Phi\rightarrow e^{2i\alpha}\Phi \quad {\rm or} \quad \Phi\rightarrow e^{i\alpha}\Phi \,,
\end{align}
respectively.  
Notice that the second term is invariant under the $Z_2$-symmetry 
\begin{align}  \label{Z2Phi}
 \qquad \Phi\, \rightarrow - \, \Phi \,.
\end{align}
This may lead to the creation of domain walls, which might be problematic and will be discussed later.    

In the above DFSZ type model, the axion today is mostly represented by the phase of the PQ field $\Phi$ with small admixtures of phases from the two Higgs doublets and the quark condensate. 
 
However, during inflation the story changes. 
As in the previous case, we assume that  the $\Sigma, H_u, H_d, \Phi$-fields interact with the inflaton $S$ in such a way that during inflation all their masses become positive and large. 
At the perturbative level, both the GUT symmetry as well as the $U(1)_{PQ}$ would be restored. 
  
\begin{figure}
    \centering
    \begin{tikzpicture}
    \def\r{1}
    \begin{feynman}
        \vertex[blob] (t) at (0,0){};
        \vertex (5) at (-1.5*\r,0){$\bar{\mathbf{5}}$};
        \vertex (101) at (0.75*\r,1.5*0.866*\r){$\mathbf{10}$};
        \vertex (102) at (0.75*\r,-1.5*0.866*\r){$\mathbf{10}$};
        \vertex (103) at (1.5*\r,0){$\mathbf{10}$};

        \diagram{
            (5) --[fermion] (t),
            (101) --[fermion] (t),
            (102) --[fermion] (t),
            (103) --[fermion] (t)};
    \end{feynman}
    \end{tikzpicture}
    \caption{A diagrammatic representation of the 't Hooft vertex in the minimal $SU(5)$ theory. The fermion in the anti-fundamental $\bar{\mathbf{5}}$ representation contributes one zero mode and the fermion in the antisymmetric $\mathbf{10}$ representation contributes three-zero modes, leading to the form \eqref{detTU5}.}
    \label{fig:minimalthooft}
\end{figure}
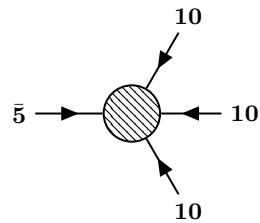

However, non-perturbatively we must take into account the fermion condensate, which forms due to strong coupling effects.       
The PQ symmetry is then broken by the condensate of the 't Hooft determinant.  
For a single generation, this determinant has the form (see Fig. \ref{fig:minimalthooft}),
\begin{align} \label{detTU5}
    \frac{1}{\Lambda_C^2} (\mathbf{10}_{jk}\mathbf{10}_{mn}\mathbf{10}_{il}\mathbf{\Bar{5}}^l) \,\epsilon^{ijkmn}.
\end{align}
The phase of the corresponding condensate, $\eta_{\rm F}$, is the axion.
It is remarkable that this degree of freedom literally represents the early Universe counterpart of the $\eta'$-meson of QCD. 
Indeed, as pointed out in \cite{Dvali:2005an} (see also, 
\cite{Dvali:2013cpa, Dvali:2017mpy, Dvali:2022fdv, Dvali:2024zpc, Dvali:2025pcx, Dvali:2025xur}) the $\eta'$-meson would play the role of an exact quality axion in case of zero quark masses, as it would fully cancel the $\bar{\theta}$-term. 
This is precisely what happens in the early confining epoch discussed above.

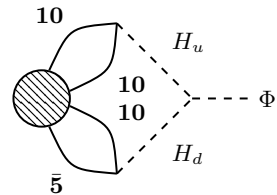
\begin{figure}
    \centering
    \begin{tikzpicture}
    \def\r{1}
    \begin{feynman}
        \vertex[blob] (t) at (0,0){};
        \vertex (x1) at (\r,\r);
        \vertex (x2) at (\r,-\r);
        \vertex (P) at (2*\r,0);
        \vertex (P2) at (3*\r,0){$\Phi$};

        \diagram{
            (x1) --[solid, half right, out = 30, in = 150, edge label = \(\mathbf{10}\) ] (t) --[solid, half right, in = 150, out = 30, edge label = \(\mathbf{10}\)] (x1),
            (x2) --[solid, half right, out = 30, in = 150, edge label = \(\bar{\mathbf{5}}\)] (t) --[solid, half right,  out = 30, in = 150, edge label = \(\mathbf{10}\)] (x2),
            (x1) --[scalar, edge label = \(\ H_u\)] (P),
            (x2) --[scalar, edge label' = \(\ H_d\),] (P),
            (P) --[scalar] (P2)};
    \end{feynman}
\end{tikzpicture}
    \caption{A diagram that gives rise to the tadpole contribution for $\Phi$.}
    \label{fig:dfsztadpole}
\end{figure}

Notice that in case of the second term in \eqref{HuHdPP}, even after taking into account the virtual Higgs effects, the condensate does not generate a tadpole for the PQ field. 
This is due to the protective $Z_2$ symmetry \eqref{Z2Phi}. 
In contrast, in the case of the first term in \eqref{HuHdPP}, the following tadpole is generated (due to the diagram in Fig. \ref{fig:dfsztadpole}), 
\begin{align} 
  \mu_{\Phi} g^ug^d \Lambda_C^2\Phi , 
\end{align} 
which induces the VEV, 
\begin{align} 
  \langle \Phi \rangle \sim \frac{1}{M^2_\Phi(S)} \mu_{\Phi} g^ug^d \Lambda_C^2  \sim   \frac{1}{M_{GUT}^2} \mu_{\Phi} g^ug^d \Lambda_C^2. 
\end{align}   
As a result, the axion, which is mainly represented by the phase of the fermion condensate $\eta_F$, acquires a small admixture, $\sim \langle \Phi \rangle/\Lambda_c$, from the phase $\eta_{\Phi}$ of the PQ field.
 
\section{The Gauge Axion} 
We now wish to discuss our scenario within the gauge axion realization, which was originally introduced in \cite{Dvali:2005an}. 
In this realization, the axion is introduced as an intrinsic part of the QCD gauge redundancy, without reference to any anomalous global symmetry.   
In the present context, the gauge group  is $SU(5)$ with the gauge field matrix $A_{\mu} \equiv A_{\mu}^aT^a$, where $T^a, ~a=1,2,...,24$ are the generators of $SU(5)$. 
 
Following \cite{Dvali:2005an}, the axion is introduced as the anti-symmetric  $2$-form field $B_{\mu \nu}$.
This field differs from an ordinary Kalb-Ramond field in the sense that under the $SU(5)$ gauge redundancy, 
\begin{align} \label{gaugeU}
    A_{\mu} \rightarrow U A_\mu U^{\dagger} + U\partial_{\mu}{U^{\dagger}}, ~ {\rm  with} ~ U \equiv \exp(iT^C\omega^C)\,,
\end{align}  
the $2$-form shifts as, 
\begin{align}\label{shift}
    B_{\mu\nu} \, \rightarrow \, B_{\mu\nu} \, + \, \frac{1}{f_a}\Omega_{\mu\nu}\,,
\end{align}
where $f_a$ is the axion scale. 
At the level of the low energy EFT, the scale $f_a$ represents a free parameter. 
However, it also serves as the upper bound on the cut-off of the EFT (see below). 
 
Notice that under the gauge transformation (\ref{gaugeU}), the $SU(5)$ Chern-Simons  (CS) $3$-form,  
\begin{align} \label{CS}
    C_{\mu\nu\alpha}  \equiv {\rm tr} \left(A_{[\mu}\partial_{\nu}A_{\alpha]} + \frac{2}{3}A_{[\mu}A_{\nu}A_{\alpha]}\right) \,,
\end{align}
shifts as, 
\begin{align} \label{ShiftQCD}
    C_{\mu\nu\alpha} \rightarrow C_{\mu\nu\alpha} + \partial_{[\mu}\Omega_{\nu\alpha]}\,,
\end{align}  
with $\Omega_{\mu\nu} \equiv A_{[\mu}^a\partial_{\nu]}\omega^a$. 
Correspondingly, the gauge axion $B_{\mu\nu}$ acts as a St\"uckelberg field for $C_{\mu\nu\alpha}$ \cite{Dvali:2005an}.   
This feature solves the strong-$CP$ problem exactly to all orders in the operator expansion. 

Indeed, the structure of the theory is fixed by the gauge invariance,
since $B_{\mu \nu}$ can enter the action only via the following gauge invariant combination with the CS $3$-form: 
\begin{align}\label{eq:Stuckelberg}
    \tilde{C}_{\mu\nu\alpha} \equiv  C_{\mu\nu\alpha} - f\partial_{[\mu}B_{\nu\alpha]}\,.
\end{align}
The lowest order invariant is 
\begin{align} \label{MtermC}
    \frac{1}{f^2}\tilde C_{\mu\nu\lambda}\tilde C^{\mu\nu\lambda}\,.
\end{align}
One can therefore add arbitrary higher order invariants without affecting the vacuum structure. 

It is straightforward to see \cite{Dvali:2005an, Dvali:2013cpa, Sakhelashvili:2021eid, Dvali:2022fdv, Dvali:2023llt} that the above theory is dual to the axion $a$, with an exact $CP$-conserving minimum, in which $G\tilde{G} =0$ is strictly satisfied, implying $\bar{\theta} =0$.  
This feature is maintained under arbitrary continuous deformations of the theory to all orders in the operator expansion \cite{Dvali:2005an}. 
Thus, from the point of view of the vacuum structure, the gauge axion is equivalent to the PQ one, but with exact quality imposed by the underlying gauge symmetry.  
This feature also creates a fundamental difference in the UV. 
The gauge axion cannot be UV-completed into a complex scalar. 
  
Embedding the gauge axion theory into our cosmological setup is straightforward.
No extra fermions or Higgs fields are required.  
Likewise, there is no PQ field $\Phi$ involved. 
The only new term is the gauge invariant eqn. \eqref{MtermC}, plus an arbitrary series of higher dimensional gauge invariants.    
The above introduces a fundamental difference between the gauge axion and KSVZ or DFSZ type scenarios.  
For our purposes the important fact is that there exists no phase of a linearly realized PQ symmetry. 
Instead, the theory is always in the St\"uckelberg phase of the gauge symmetry.  
This applies equally to the inflationary epoch as well as to any other epoch describable within the EFT. 

Nevertheless, in order to dilute the axion field during inflation, the mechanism of achieving a period of a strong gauge coupling, as discussed before, is still operative.  However, the axion decay constant, $f_a$, represents a fundamental scale of the theory and is independent of the inflaton field.  

\section{The Axion Composition Mode}

As already pointed out, an intrinsic property of the discussed KSVZ and DFSZ scenarios is that the axion composition during inflation differs from its composition today.
Let us take into account the cosmological effect of this change. 
For simplicity, we discuss the KSVZ case, ignoring the ordinary fermions. 
We focus on the regime in which the mass of the PQ field $\Phi$ is positive during inflation. 
In such a case, the spontaneous breaking of the PQ symmetry is predominantly due to the fermion condensate $\langle \Bar{\Psi}^r{\Psi}_r \rangle $, whereas a small VEV of $\Phi$ can be sourced by the fermion condensate \eqref{PhiVEV}.  
Correspondingly, the axion is predominantly made up of $\eta_F$, which corresponds to the phase of the condensate.  
The phase of $\Phi$, which we denote by $\eta_{\Phi}$, contributes very little.  
To be more precise, the genuine axion, $a$, represents the following superposition (for more details see the Appendix \ref{sec:AppB}) 
\begin{align} \label{EtaXi}
    a = \eta_F \cos(\xi) - \eta_{\Phi} \sin(\xi)\,,
\end{align}
where the angle $\xi$ is given by 
\begin{align} \label{tanXi}
    \tan (\xi) = \frac{\langle \Phi \rangle}{\langle \Bar{\Psi}^r\Psi_r\rangle^{1/3}} \,.
\end{align}
The important thing is that $\xi$ changes over time. 
In particular, during the inflationary strong QCD phase it is miniscule, $\xi \sim (\Lambda_C/M_{GUT})^2  \sim 10^{-20}$, whereas in today's universe it is $\xi \simeq \pi/2$. 

Correspondingly, $\xi$ is described by a time-dependent degree of freedom which 
carries energy. 
However, this degree of freedom is orthogonal to the axion field and takes no part in the misalignment of the $\theta_{\rm eff}$. 
In today's universe, its mass is much higher than the one of the axion and carries no significant energy density.  

Now, the time-dependence of $\xi$ can certainly produce some axion gas, which however is unrelated with the coherent mode parameterizing the misalignment of $\theta_{\rm eff}$ which we are interested in. 
The production of axion radiation depends on the details of reheating and must be investigated on a model-dependent basis.  
However, it is typically expected to consist of axions of energies comparable to the reheating temperature. 

\section{Cosmic Strings, Domain Walls and Magnetic Monopoles} 

We shall now discuss the spectrum of topological defects in our scenario.  
These include both axionic string-wall systems and magnetic monopoles. 
  
\subsection{Axionic String and Domain Walls} 
We start by discussing axionic defects. 
Traditionally, these represent the axionic cosmic strings with domain walls attached to them (for various aspects, see e.g. \cite{Vilenkin:1982ks, Lazarides:1982tw, Lazarides:1984zq}). 

An intrinsic feature of our scenario is that the axion exists as a valid degree of freedom continuously, starting from a certain moment during inflation all the way to the present epoch. 
In the gauge axion scenario, this condition is automatically fulfilled during the period of the universe's history, including the entirety of inflation, that fits within the validity of  the EFT description. 
  
For KSVZ or DFSZ type scenarios, this  is not automatic and requires a continuity of the phase with spontaneously broken $U(1)_{PQ}$-symmetry.   
However, the order parameter responsible for this breaking can change in time, alternating between the fermionic condensate and the PQ field $\Phi$.     
In particular, as discussed above, during an early strong coupling epoch, the PQ symmetry can be predominantly broken by the condensate of the fermionic t' Hooft determinant, in both KSVZ or DFSZ type realizations. 
For the fermion condensate dominated scenario, string-wall systems form together with the fermion condensate.   
These string-wall systems of the quark condensate were discussed in detail in \cite{Dvali:2025xur}. 
As shown there, the quark condensate 
significantly affects the features of the axionic strings such as the structure of the fermionic zero modes.
    
As was recently argued in \cite{Dvali:2025ivw}, strings sourcing the gauge axion are fundamental, i.e., not describable as smooth solitons within the EFT.   
This finding naturally fits the understanding that a UV-completion of the gauge axion requires going beyond the EFT cutoff $f_a$.  
 
The generic expectation of our early strong gauge coupling scenario is that no post-inflationary formation of axionic string-wall systems takes place.
The defects are either pre-existing or form during inflation. 
Hence, they are all inflated away.
After inflation, the universe is left with a homogeneous axion field, which can be a viable dark matter candidate for an arbitrarily large value of $f_a$.  
     
\subsection{Magnetic Monopoles} 
   
Another type of topological defects are magnetic monopoles. 
In our scenario the GUT symmetry can be restored during certain periods of inflation. 
If the transition breaking the GUT symmetry takes place while the universe is still inflating, monopoles are inflated away and can be cosmologically irrelevant \cite{Guth:1980zm}.  
On the other hand, if the GUT symmetry breaking happens towards the end or after inflation, monopoles can potentially be produced in abundance.
In such a case, one needs to make sure that the surviving number will not cause a cosmological monopole problem \cite{Preskill:1979zi, Zeldovich:1978wj}.
    
The number density of monopoles depends on factors such as the remaining number of e-foldings until the end of inflation and the reheating temperature. 
In the evaluation of these factors, we first notice that a high reheating temperature by no means implies the monopole production, since for a large potion of the parameter space the GUT symmetry is not restored at arbitrarily high temperatures \cite{Dvali:1995cj}.  
     
Next, even if monopoles are produced after inflation, their number can be drastically reduced by at least two mechanisms.    
The first one, introduced by Langacker and Pi~\cite{Langacker:1980kd}, proposed that for a finite time interval monopoles could get connected by strings and thus annihilate.
     
As a more minimalistic approach, note that remarkably, the $SU(5)$ GUT possesses a built-in mechanism for the ``erasure"  of monopoles by unstable domain walls formed in the same phase transition \cite{Dvali:1997sa, Brush:2015vda, Bachmaier:2023zmq}.    
The key idea behind the erasure mechanism is that, if the coefficient $\mu_{\Sigma}$ of the cubic self-interaction term in \eqref{scalarpot} is relatively small, the theory exhibits an 
approximate discrete symmetry $\Sigma \rightarrow - \Sigma$.  
Due to this, the very same transition that forms monopoles also forms unstable domain walls. 
These walls collapse soon after their formation and in this process they get rid of the monopoles.  
The reason for this is that, when a monopole encounters a wall, it unwinds and disappears.  
In this way, the unstable walls sweep away the monopole problem.

Finally, we would like to remark that our cosmological scenario brings some new fundamental questions about monopole production that have not been discussed previously.   
In particular, the breaking (Higgsing) of the GUT symmetry can happen immediately after the confining phase. 
In this phase the monopoles are condensed and correspondingly, the magnetic charges are screened.  
It is an interesting question how an immediate sequence of confining and Higgs phases affects the dynamics of monopole formation. 

\section{Beyond our analysis} 

We would like to make some comments about possible extensions of this study beyond the presented analysis.
We wish to note that we have ignored the relative variation of $\bar{\theta}$ due to a possible temperature dependence of the phase of the fermion mass determinant. 
This can create an additional misalignment angle $\theta_{\rm eff}$ between the present and inflationary epochs. 
However, as already pointed out in \cite{Koutsangelas:2022lte}, the effect is negligible. 
This is due to the extremely suppressed renormalization of the phase of the quark determinant, which at zero temperature amounts to a change $\delta \bar{\theta} \sim 10^{-16}$ \cite{Ellis:1978hq}.  
In the present context, additional misalignment does not come from $\delta \bar{\theta}$, which is absorbed in todays axion minimum, but by its relative change in the early epoch, which is a higher order effect.   
  
Treating the minimal $SU(5)$ as the most well-known prototype, we have ignored its usual potential inconsistencies and challenges. 
For example, such are wrong relations between the fermion masses as well as threshold effects at the unification scale. 
As customary, these must be taken care of by higher order operators and/or additional fields.
None of this is expected to qualitatively affect our analysis.
If anything, the need for higher dimensional operators only makes the motivation for the couplings \eqref{GaugeKin} stronger. 

Our scenario can easily be extended to a larger grand unified group such as $SO(10)$. 
From the point of view of increasing the QCD scale, $\Lambda_C$, in the restored-symmetry epoch, such embeddings would only be beneficial. This is due to the fact that for larger groups the contribution of the gauge fields in the beta function gets larger. 

In our discussion, we assumed that the axion has an  exact quality. 
It has been argued that this is a necessary requirement for a consistent embedding of QCD 
(or any other gauge sector with a $\theta$-vacuum structure) in gravity \cite{Dvali:2018dce, Dvali:2022fdv, Dvali:2023llt, Dvali:2024dlb, Dvali:2024zpc}.                        
This requirement is automatically satisfied by the gauge axion \cite{Dvali:2005an}. 
For PQ realizations, it implies that the global symmetry is broken exclusively by the gauge anomaly.  This is an additional condition which we have assumed to be satisfied. 
Compromising the axion quality can result in additional cosmological contributions to the misalignment angle (see, e.g., \cite{Michael}). 

However, we must stress that our philosophy disfavors any artificial explicit breaking of the axion shift symmetry, beyond the gauge anomaly of QCD (or its extensions). 

\section{Outlook}  

In this paper we have discussed the axion's early relaxation scenario \cite{Dvali:1995ce}, which removes the standard cosmological upper bound \eqref{AxionBound} on the axion scale. 
This is due to the fact that the gauge coupling could easily become strong in a very early cosmological epoch, leading to the generation of a large axion mass and the subsequent dilution of the axion energy an early period of coherent oscillations. 
In the analyses \cite{Preskill:1982cy, Dine:1982ah, Abbott:1982af} leading to the bound 
\eqref{AxionBound}, this regime is not taken into account, implicitly assuming that the the axion only learns about its vacuum after the most recent QCD phase transition.    
As argued in \cite{Dvali:1995ce} and confirmed by our analysis, this is a rather strong assumption not accommodated by many scenarios.  

The effect of early relaxation can be especially prominent within the inflationary cosmology, due to strong field-dependent variations of the gauge coupling. 
Such displacements of the fields and their effective couplings/masses is the very essence of the inflationary cosmology.  
It is therefore highly unlikely that the QCD gauge coupling was not subjected to variations during the inflationary epoch.  
  
In the present work we gave an explicit realization of this proposal within inflationary cosmology, implemented in the framework of grand unification. 
We have pointed out that inflation gives at least two separate avenues to strengthen the gauge coupling in the early epoch. 
     
One way to achieve this is through a direct inflaton dependence of the gauge coupling, imposed by generic gauge invariant operators that couple the inflaton to the gauge fields. 
Not only are such operators are permitted by symmetries, but often they are a direct consequence of embedding the gauge sector within the inflationary cosmology. 

The mechanism of early relaxation can be implemented already at the level of the SM coupled to an inflaton.   
However, grand unification provides a separate mechanism to generate early strong coupling dynamics and confinement. 
This takes place in regimes in which the GUT symmetry is restored during some period of inflation. 
In this case, the gauge coupling of $SU(5)$ can become strong due to the RG running. 

We discussed the above cosmological scenario for three different realizations of the invisible axion idea: the KSVZ  \cite{Kim:1979if, Shifman:1979if},  DFSZ \cite{Zhitnitsky:1980tq, Dine:1981rt} and gauge axion  \cite{Dvali:2005an} scenarios. 
   
We have shown that in the strong gauge coupling epoch the axion is a valid degree of freedom in all three theories.  
In the gauge axion scenario \cite{Dvali:2005an}, this is automatic, since the Chern-Simons gauge symmetry is in the Higgs phase throughout the regime of validity of the EFT.  

However, the presence of the axion field in the strong coupling epoch is equally true in KSVZ and DFSZ scenarios, albeit in a more profound way.
In these scenarios, in the present day, the PQ-symmetry is spontaneously broken by the VEV of a singlet $\Phi$ and the axion originates from its phase.  
In an early epoch however, the VEV of the PQ field $\Phi$ could vanish, due to couplings to the inflaton. 
Nevertheless, even in such a case, the axion remains a well-defined entity, since it originates from the phase of the fermionic 't Hooft determinant, which is generated by strong coupling dynamics. 
This is an exact statement which, in particular, follows from arguments based on the topological susceptibility of the vacuum \cite{Dvali:2017mpy}. 

In particular, if the strong coupling period coincides with a restored phase of the GUT symmetry, the VEV of the 't Hooft determinant is generated by the entirety of the $SU(5)$ instantons.     
This ensures the continuity of the ``cosmological memory" of the axion potential, reducing the misalignment of $\theta_{\rm eff}$ through coherent oscillations in the early epoch.  
    
We have also discussed the fate of topological defects, such as axionic string-wall systems as well as GUT magnetic monopoles in three scenarios and concluded that they do not pose any significant problem or a constraint.
If the epoch of strong gauge coupling takes place sufficiently close towards the end of inflation, some gravitational wave-signatures of string-wall systems might potentially be of observational interest.
However, due to the wideness of the phenomenologically acceptable parameter space, concrete predictions correlating gravitational wave signatures with axion dark matter abundance require a narrowing of the class of viable models by additional criteria. 

Finally, we would like to comment that recently, it has been argued \cite{Dvali:2024zpc, Dvali:2025pcx} that various consistency requirements indicate the existence of a new particle, the so-called $\eta_{\rm w}$-meson, which plays an analogous role for the anomalous $B+L$-symmetry of the SM to the one assumed by the axion or the $\eta'$-meson for the chiral symmetry of QCD.  
In the current Universe, this degree of freedom, at least partially, resides in the phase of the 't Hooft determinant generated by the $SU(2)$ instantons. 
Since the electroweak gauge theory is in the Higgs phase, the $\eta_{\rm w}$-meson is extremely light and carries no significant energy.    
However, the story could have been very different in the early Universe,  due to early relaxation experienced by the $\eta_{\rm w}$-meson.  
In particular, during inflation, the $SU(2)$-theory can experience a strongly-coupled confining phase, generating a large mass for the $\eta_{\rm w}$-meson. 
The resulting cosmological implications remain to be explored.

{\textsl{\bf Acknowledgments}}\;---\;

We thank Otari Sakhelashvili for valuable discussions.
The Feynman diagrams in this work were created using Tikz-Feynman \cite{ELLIS2017103}.
 
This work was supported in part by the Humboldt Foundation under the Humboldt Professorship Award, by the European Research Council Gravities Horizon Grant AO number: 850 173-6, by the Deutsche Forschungsgemeinschaft (DFG, German Research Foundation) under Germany's Excellence Strategy - EXC-2111 - 390814868, and Germany's Excellence Strategy under Excellence Cluster Origins. \\
 
Disclaimer: Funded by the European Union. 
Views and opinions expressed are however those of the authors only and do not necessarily reflect those of the European Union or European Research Council.
Neither the European Union nor the granting authority can be held responsible for them.

\appendix
\section{The Confinement Scale of $SU(5)$}
\label{AppA}

In this appendix we elaborate on the strong coupling scale of  $SU(5)$ during inflation.  
We focus on the inflationary epoch, in which the GUT Higgs fields, get large positive masses and zero VEVs. 
In case of the KSVZ and DFSZ realizations, we also assume that the VEV of the the PQ scalar $\Phi$ is zero as well. 
In the case of the gauge axion, there is no such field and the axion exists as a fundamental degree of freedom.

Thus, in all realizations of the axion, in the epoch of our interest, below the scale $M_{GUT}$, the perturbative  spectrum of the theory consists of massless gauge fields and massless fermions.   
In this setup we wish to estimate at what scale, $\Lambda_C$,  the $SU(5)$ gauge theory can hit the infrared strong coupling regime via RG running. 
  
We shall preemptively assume that the Hubble parameter 
during this epoch is significantly lower than this scale, ${\mathcal H} \ll \Lambda_C$.
This allows us to safely ignore the effects of space-time curvature on the RG evolution. 
  
Next, we need to choose the boundary conditions for the $SU(5)$ gauge couplings at the GUT scale. 
This requires some clarification, since we are effectively 
studying the theory in a ``vacuum" which is drastically different from the current one. 
Correspondingly, the value of the $SU(5)$ gauge coupling  evaluated  via a bottom-up running from the current vacuum is in general very different.   
The right EFT way of thinking about the situation is that we are dealing with a theory that has a landscape of vacuum states, parameterized by the master mode $S$. 
The spectrum of the theory, and correspondingly the outcome of the RG running, depend on $S$. 
   
In the full theory, of course, one can ``rewind" the inflaton clock and extrapolate the unified value of the coupling obtained in today's vacuum to the one in the inflationary epoch. 
However, in the present context, this is unnecessary due to a number of reasons. 
First,  as pointed out, we are dealing with a simple prototype rather then a phenomenologically fully consistent
$SU(5)$, which would require certain additional problems to be fixed. 
   
Therefore, in the present case, as a proof of concept, it is fully sufficient to illustrate that, starting from a weak gauge coupling at the GUT scale in un-Higgsed $SU(5)$, the strong coupling scale reached by RG running  can be high.  
  
For illustrative purposes, as a reasonable boundary condition we choose \cite{Koutsangelas:2022lte},
\begin{align}
    \alpha_S(M_{\rm GUT}) \simeq 0.0224\ .
\end{align}
Since the Higgs scalars are heavy and do not contribute, we obtain the confinement scale of unbroken $SU(5)$ from the RG equation as
\begin{align}
\begin{split}
    \Lambda_C &= M_\mathrm{GUT}\exp{\left\{-\frac{2\pi}{\beta_0}\left(\frac{1}{\alpha_{SU(5)}(M_\mathrm{GUT})} - 1\right)\right\}} \\
    &\sim 10^{16}\mathrm{GeV}\exp{\left\{-\frac{2\pi}{\beta_0}\left(\frac{1}{0.0224} - 1\right)\right\}}\ ,
\end{split}
\end{align}
where $\beta_0$ is model dependent.
In the DFSZ case, 
only the SM fermions contribute and
\begin{align}
    \beta_0 = \frac{11}{3}\times 5 - \frac{2}{3}\left[\frac{3}{2} + \frac{1}{2}\right]\times 3 = \frac{43}{3} .
\end{align}

For the KSVZ model we add the PQ fermions as 2 left-handed Weyl fermions in the fundamental and anti-fundamental representation, respectively. 
Since the VEV of $\Phi$, eqn. \eqref{PhiVEV}, induced from the tadpole of the 't Hooft determinant is highly suppressed, the $\Bar{\Psi}^i,\Psi_{i}$ fermions can be considered massless.
This determines $\beta_0$ to be
\begin{align}
    \beta_0 = \frac{11}{3}\times 5 - \frac{2}{3}\left[\frac{3}{2} + \frac{1}{2} \right]\times 3 - \frac{2}{3}\left[\frac{1}{2} + \frac{1}{2}\right]= \frac{41}{3}.
\end{align}
Hence, we obtain for the confinement scales of the two models:
\begin{align}
    \mathrm{DFSZ:}\quad \Lambda_C \sim 5\times10^7\mathrm{GeV}\
\end{align}
and
\begin{align}
    \mathrm{KSVZ:}\quad \Lambda_C \sim 2\times10^7\mathrm{GeV}.
\end{align}
In the gauge axion scenario only the SM fermions contribute to the $\beta$-function, just as in the DFSZ case, such that we also obtain $\Lambda_C \sim 5\times 10^7$GeV.

\section{The Composition of the Axion} 
\label{sec:AppB}
It is useful to write the two order parameters that break the $U(1)_{PQ}$-symmetry in terms of a modulus and the phases of the respective degrees of freedom 
\begin{align} \label{PhiPsiPsi}
    \Phi \equiv \rho_{\Phi} e^{i \frac{\eta_{\Phi}}{\rho_{\Phi}}},~ 
    \langle \Bar{\Psi}^i\Psi_{i} \rangle \equiv 
    \Lambda_C^2 \rho_{F} e^{i \frac{\eta_{F}}{\rho_{F}}} \,,   
\end{align}
where, as we already know, $\rho_{\Phi}$ and $\rho_{F}$ as well as $ \Lambda_C$ are time-dependent through the inflaton VEV. 
Plugging this in the coupling we get  
\begin{align}
    g^F \Phi \Bar{\Psi}^i\Psi_i  +  {\rm h.c.} = - 2g^F \Lambda_C^2 \rho_{\Phi} \rho_{F} \cos \left ( \frac{\eta_{\Phi}}{\rho_{\Phi}}  
    +  \frac{\eta_{F}}{\rho_{F}} \right ) \,,
\end{align}             
where the phase of $g^F$ has been absorbed by a shift of the phases.   
This shows that the mode, 
\begin{align}
  \eta_{\rm heavy}  \equiv \eta_{\Phi} \cos(\xi) +  \eta_{F} \sin(\xi),\ 
  \ {\rm with } \ \tan{\xi} \equiv \frac{\rho_{\Phi}}{\rho_{F}}\,, 
\end{align} 
gets the mass $M^2_{\rm heavy} = 4g^F \Lambda_C^2 (\sin(2\xi))^{-1}$. 
The orthogonal mode,
\begin{align} \label{TRUEa} 
   a \equiv   \eta_{F} \cos(\xi) -  \eta_{\Phi} \sin(\xi)\,, 
\end{align} 
is the axion, which recovers eqn. \eqref{EtaXi}. 
The axion gets its mass,
\begin{align}
   m_a^2  =  
    \frac{\Lambda_C^3}{\rho_{F}}  \cos^2(\xi)\,,  
\end{align} 
from the instantons. 
During the inflationary period of our interest, this mass is of order  $m_a \sim  \Lambda_C $ in full accordance with the previous discussions in the main text. 
This leads to the early relaxation of $\theta_{\rm eff}$. 
  
Let us now focus on the degree of freedom which is responsible for the time-dependence of the scales $\Lambda_C, \rho_{\Phi}, \rho_F$ and correspondingly of the angle $\xi$.  
For convenience, we shall refer to it as the ``master" mode.
This mode resides in the fluctuations of the absolute values $\rho_{\Phi}$ and $\rho_F$ and its mass during inflation is set by the curvature of the inflaton potential. 

In order to see this, let us complete the above simplified example into an inflationary model by coupling the PQ field  $\Phi$ to an inflaton $S$. 
For simplicity, we disregard the GUT scalar fields and the SM fermions. 
The scalar potential and Yukawa terms read:  
\begin{align}
   \frac{1}{2}\lambda_{\Phi} \left (|\Phi|^2 - \frac{M_{\Phi}^2}{\lambda_{\Phi}}\right )^2  +  \kappa_{\Phi} S^2 |\Phi|^2  + 
    g^F \Phi \Bar{\Psi}^i\Psi_i  +  {\rm h.c.}\,. 
\end{align} 
For $|S| \gg  \frac{M_{\Phi}}{\sqrt{\lambda_{\Phi}}}$, the perturbative minimum of the potential is achieved at $\Phi=0$, and the potential is flat in the $S$ direction with the constant value $V=  \frac{1}{2} \frac{M_{\Phi}^4}{\lambda_{\Phi}}$. 
However, since the gauge sector enters the strong coupling regime, the potential gets corrected. 
Substituting \eqref{PhiPsiPsi} and taking into account the instanton corrections, we get the following scalar potential  
\begin{align}
\begin{split}
    V_{\rm eff} &=  \frac{1}{2}\lambda_{\Phi} \left(\rho_{\Phi} ^2 - \frac{M_{\Phi}^2}{\lambda_{\Phi}}\right )^2 + \kappa_{\Phi} S^2 \rho_{\Phi} ^2   
    \\ &-2g^F \Lambda_C^2 \rho_{\Phi} \rho_{F} \cos \left( \frac{\eta_{\Phi}}{\rho_{\Phi}}  +  \frac{\eta_{F}}{\rho_{F}} \right ) \\ 
    &-\Lambda_C^3  \rho_{F} \cos \left ( \frac{\eta_{F}}{\rho_{F}} - \bar{\theta} \right )\,, 
\end{split}
\end{align}     
where the second and the third lines come from the substitution of the fermion condensate into the Yukawa coupling and into the 't Hooft determinant, respectively.  
We must keep in mind that the QCD scale  $\Lambda_C(S)$ is a function of $S$ and that $\rho_{F} \sim \Lambda_C(S)$. 
  
The non-perturbative contribution corrects the flatness of the inflaton potential.  
For $|S| \gg  \frac{M_{\Phi}}{\sqrt{\lambda_{\Phi}}}$, the correction to the potential to leading order is 
\begin{align}
      \Delta V = -\frac{1}{S^2}g_F^2\ \Lambda^4_C(S)\ \rho^2_F(S)
      \sim \frac{1}{S^2}\Lambda_C^6(S)\,.
\end{align}
 
It is clear that regardless the value of $S$, the minimum of the last term is achieved at $\frac{\eta_{F}}{\rho_{F}} = \bar{\theta}$, implying $\theta_{\rm eff} =0$. 
Simultaneously, we have $\frac{\eta_{\Phi}}{\rho_{\Phi}}  =  -\frac{\eta_{F}}{\rho_{F}}$, which minimizes the Yukawa coupling term. 
Thus, $\theta_{\rm eff}$ is relaxed to zero at any moment of the strong coupling epoch.
What changes in time is the composition of the axion field, which is given by \eqref{TRUEa}. 
    
The role of the master mode determining this identity of the axion field in the inflationary epoch is played by $S$.  Through $S$ the parameters $\Lambda_C, \rho_{\Phi}, \rho_F$ and $\xi$ are time-dependent.  

The time-dependence of these parameters leads to the time-dependence of the axion mass as well as of its decay constant, without affecting the minimum of its potential. 
Correspondingly, the relaxation takes place towards $\theta_{\rm eff} = 0$ in every epoch, regardless of the master mode.  
 
In today's universe the master mode is mainly represented by the modulus of the $\Phi$ field and has a very high mass, of the order of today's value of the axion decay constant  $M_{\Phi} \sim \sqrt{\lambda_{\Phi}} f_a$.  
Its fluctuations can therefore safely be ignored. 
 
However, the time-dependence of the axion mass through variations of the master mode will in general result in particle creation. 
The produced axions are not contributing into the coherent mode that misaligns $\theta_{\rm eff}$ but rather represent a gas of axions. 
The most violent production is expected to take place during reheating. 
In the above prototype model this happens after the inflaton field drops below the critical value  $|S| = \frac{M_{\Phi}}{\sqrt{\lambda_{\Phi}}}$ and the PQ field becomes unstable and relaxes towards today's vacuum. 
This is not much different from a possible thermal production of axions during the PQ phase transition in ordinary scenarios and is not specific to the present study. 

\bibliographystyle{utphys}
\bibliography{bibliography.bib}		

\end{document}